\documentclass[12pt]{iopart}

\usepackage{graphicx,bbm}

\usepackage{iopams}
\newcommand{\ket}[1]{|#1\rangle}
\newcommand{\bra}[1]{\langle#1|}

\newcommand{\proj}[1]{\ket{#1}\!\bra{#1}}
\newcommand{\ot}[0]{\otimes}

\begin{document}

\title[Checking the optimality of entanglement witnesses: an application to
$\ldots$]{Checking the optimality of
entanglement witnesses: an application to 
structural physical approximations}

\author{R Augusiak$^1$, J Bae$^2$, J Tura$^1$, M Lewenstein$^{1,3}$}
\address{$^1$ ICFO--Institut de Ciencies Fotoniques, Mediterranean Technology
Park, 08860
Castelldefels (Barcelona), Spain}
\address{$^2$ Center for Quantum Technologies,
National University of Singapore, 3 Science Drive 2, Singapore 117543}
\address{$^3$ ICREA--Institucio Catalana de Recerca i
Estudis Avan\c{c}ats, Lluis Companys 23, 08010 Barcelona, Spain}

\begin{abstract}
In 2008, the conjecture that structural physical
approximations to optimal entanglement witnesses are separable states (in
general unnormalized) was posed. In an attempt to disprove it, in (Ha K-C and
Kye S-H, \textit{Separable states with unique decompositions},
arXiv:1210.1088v3), Ha and Kye proposed a decomposable entanglement witness
whose SPA is entangled and argued that it is optimal. In this note, which is
based on a comment to the latter work (Augusiak \textit{et al},
\textit{Comment on ``Separable states with unique decompositions''},
arXiv:1304.2040v1), we show, both analytically and numerically, that this
entanglement witness is not optimal, and as such it is not a counterexample to
the conjecture. To this end, we make use of a method for checking optimality of
entanglement witnesses developed already in (Lewenstein \textit{et al} 2000
\textit{Phys. Rev. A} \textbf{62} 052310), however, hardly exploited so far in
the literature.
\end{abstract}

\maketitle

\section{Introduction and preliminaries}
\label{intro}

Several years ago, in Ref. \cite{Korbicz} some of us conjectured that a
structural physical approximation (SPA) of an optimal positive map is
entanglement breaking, or, equivalently, that SPA of an optimal entanglement
witness is a separable state (in general unnormalized). Later, the conjecture
was supported by numerous examples of witnesses (see, e.g., Refs.
\cite{examples1,examples2,examples3,we} and also Ref. \cite{Myrheim}), both
decomposable and indecomposable, and even studied in the continuous variables
case \cite{we}. Quite surprisingly, in the indecomposable case it was recently
disproved by Ha and Kye \cite{HaKye1} (see also Ref. \cite{Storm2}), who found
an example of an optimal indecomposable entanglement witness, whose structural
physical approximation is entangled. Still, the conjecture remained unresolved
in the decomposable case and later in an attempt to elucidate this, the same
authors proposed in Ref. \cite{paper} a decomposable witness whose SPA is
entangled and claimed it to be optimal. Had the latter been true, the conjecture
would have been fully disproved. 

The main aim of this note, which is a slightly modified version of the
preprint \cite{ourpreprint} released as a comment to Ref. \cite{paper}, is to
show that the entanglement witness of Ref. \cite{paper} (see also Ref.
\cite{paper2}) is not optimal, and as such it is not a correct candidate to
disprove the conjecture. 

It should be stressed, however, that after our comment appeared on-line
\cite{ourpreprint}, the conjecture has eventually been disproved in the
decomposable case by Chru\'sci\'nski and Sarbicki \cite{Gniewko3}.
Nevertheless, we still find the optimality analysis presented in this note
valuable for researchers working in entanglement theory. First, it is in
general not an easy problem to analytically decide on optimality of an EW,
if it does not have the so-called spanning property. Second, our note presents
an instructive application of a powerful machinery for checking optimality of
EWs developed already in Ref. \cite{optimization}, but rather hardly used  in
the literature.

\paragraph{Preliminaries.} Before presenting our arguments, let us recall some
background material.
Let ${\cal H} = {\mathbb{C}}^d$ be a complex Hilbert space
of dimension $d$, $M_d$ the set of $d\times d$ matrices of complex entries,
and $\mathbbm{1}_d\in M_d$ the identity matrix. A density
operator acting on $\cal H$ is a positive semi-definite linear operator $\rho:
{\cal H}\longrightarrow {\cal H}$ of unit trace.

Consider now a finite-dimensional product Hilbert space
$\mathcal{H}_{AB}=\mathbbm{C}^{d_A}\ot\mathbbm{C}^{d_B}$ and let
$\mathcal{D}=\{\rho\in M_{d_A}\ot M_{d_B}\,|\,\rho\geq 0,\Tr\rho=1\}$
denote the set of density operators acting on $\mathcal{H}_{AB}$.
We say that $\rho\in\mathcal{D}$ is \textit{separable} if, and
only if it admits the following convex decomposition \cite{Werner}:
\begin{equation}\label{separable}
\rho=\sum_{i}p_i\rho_A^i\otimes \rho_B^i,\qquad p_i\geq 0,\qquad \sum_ip_i=1,
\end{equation}
where $\rho_X^i$ are density matrices acting
on $\mathbbm{C}^{d_X}$ $(X=A,B)$. Otherwise we say that $\rho$ is
\textit{entangled}.

Entanglement not only fundamentally distinguishes
between quantum and classical theories, but it is also a key resource for
quantum information theory (QIT) (see \cite{Hprzeglad}).
Thus, its detection in composite quantum systems is one of the central problems
in QIT (see \cite{GTprzeglad}). A crucial method for this purpose,
formulated in Ref. \cite{HorodeckiCriterion}, exploits the fact
that the set of separable states, denoted $\mathcal{D}_{\mathrm{sep}}$, is
closed and convex. Thus, it can be fully characterized by
half-spaces whose normal vectors are non-positive
semi-definite elements of the dual cone of
$\mathcal{D}_{\mathrm{sep}}$, ${\cal P}=\{W\in M_{d_A}\ot M_{d_B}\,|\,
\mathrm{Tr}(W\rho) \geq 0,  \rho \in {\cal D}_{\mathrm{sep}}\}$. Such
operators form a nonconvex set, denoted $\mathcal{W}$, and, following Ref.
\cite{Terhal}, are called \textit{entanglement witnesses}. Within this
framework, $\rho\in\mathcal{D}_{\mathrm{sep}}$ if, and only if,
$\Tr(W\rho)\geq 0$ for any $W\in \mathcal{W}$ \cite{HorodeckiCriterion}.

Clearly, not all EWs are necessary to detect all entangled states, and
the first attempt to find the minimal set of EWs, which do the job, was made in
Ref. \cite{optimization}, where the notion of optimal EW was
introduced. To recall it, let us denote
\begin{equation}\label{DW}
\Delta_W=\{\rho\in\mathcal{D}\,|\,\Tr(\rho W)<0\},
\end{equation}
and
\begin{equation}\label{PW}
\Pi_W=\{\ket{e,f}\in\mathbbm{C}^{d_A}\ot\mathbbm{C}^{d_B}\,|\,\bra{e,f}W\ket{e,f
} =0\}.
\end{equation}
Given $W_1, W_2 \in {\cal W}$, $W_1$ is called
\textit{finer} than $W_2$ if $\Delta_{W_2}\subseteq \Delta_{W_1}$. If there is
no witness finer that $W\in\mathcal{W}$, we call it \textit{optimal}
\cite{optimization}. Alternatively speaking, a given $W\in\mathcal{W}$ is
optimal iff for any $\lambda>0$ and any operator $P \geq 0$ having support
orthogonal to $\Pi_W$, $W-\lambda P \notin {\cal W}$,
i.e., the inequality
\begin{equation}\label{inequality}
\bra{e,f}W-\lambda P\ket{e,f}\geq0
\end{equation}
does not hold for some $\ket{e,f}$ (cf. Lemma 3 and
Theorem 1, Ref. \cite{optimization}). This implies a sufficient
condition for optimality: if $\mathrm{span}\Pi_W=\mathcal{H}_{AB}$, then
$W$ is optimal \cite{optimization}.

Geometrically,
$\mathrm{Ext}(\mathcal{W})\subset\mathrm{Opt}(\mathcal{W}
)\subset\partial\mathcal{W}$ (cf. \cite{Karol}), where
$\mathrm{Ext}(\mathcal{W})$ and $\mathrm{Opt}(\mathcal{W})$ denote the sets of
extremal (those generating extremal rays in $\mathcal{P}$) and
optimal EWs, respectively, while $\partial \mathcal{W}$ stands for the boundary
of $\mathcal{W}$. There exist, however,
$W\in\partial \mathcal{W}\setminus\mathrm{Opt}(\mathcal{W})$
and also $W\in\mathrm{Opt}(\mathcal{W})\setminus\mathrm{Ext}(\mathcal{W})$,
thus, in general,
$\mathrm{Opt}(\mathcal{W})\varsubsetneq \partial\mathcal{W}$ and
$\mathrm{Ext}(\mathcal{W})\varsubsetneq\mathrm{Opt}(\mathcal{W})$.
As an illustrative example of $W\in\partial {\cal W} \setminus
\mathrm{Opt}({\cal W})$ consider a line segment $W(p)=pW_{+}+(1-p)W_{-}\in
M_2\ot M_2 $ with $0\leq p\leq 1$, where
$W_{\pm}=\proj{\psi_{\pm}}^{T_B}\in\mathrm{Ext}(\mathcal{W})$
with $\ket{\psi_{\pm}}=(1/\sqrt{2})(\ket{00}\pm\ket{11})$. For any $p\neq 1/2$,
$W(p)$ is an EW, while $W(1/2)\geq 0$, and therefore $W(p)$ are not optimal
for any $p\neq 0,1$. To prove that $W(p)\in\partial\mathcal{W}$ for any $0\leq
p\leq 1$, it is enough to notice that $W(p)-\epsilon
\proj{\phi_{+}}^{T_B}\notin\mathcal{W}$ for any $\epsilon>0$ with
$\ket{\phi_+}=(1/\sqrt{2})(\ket{01}+\ket{10})$. Then,
as an example of
$W\in\mathrm{Opt}(\mathcal{W})\setminus\mathrm{Ext}(\mathcal{W})$
consider a decomposable\footnote{We call
$W\in\mathcal{W}$ decomposable if $W=P+Q^{T_B}$ with $P,Q$ being two
positive semi-definite operators and $T_B$ denoting the partial transposition
with respect to the second Hilbert space $\mathbbm{C}^{d_B}$ (the partial
transposition with respect to the first subsystem can equivalently be taken).
Otherwise $W$ is called \textit{indecomposable}. Notice that originally the
notion of decomposability was formulated for positive maps
in \cite{Storm,Wor} and later translated via the Choi-Jamio\l{}kowski
isomorphism to EWs.} witness $W=Q^{T_A}$ with $Q\in
M_2\ot M_d$ ($d\geq 2$), such that $Q\geq0$ and $\mathrm{supp}(Q)$ is a
completely entangled subspace\footnote{A subspace of
$\mathbbm{C}^{d_A}\ot\mathbbm{C}^{d_B}$ is called completely entangled if it
contains no product vectors (for more information about these objects see, e.g.,
Ref. \cite{walgate}).}. Any such EW is optimal \cite{our}, still it is not
extremal provided $\mathrm{rank}(Q)>1$. Noticeably, although
extremal or even exposed EWs form proper subsets of $\mathrm{Opt}(\mathcal{W})$,
they detect all entangled states (see Refs. \cite{Karol,HaKye,Gniewko2}).
Yet, the definition of optimal EWs is operational
in the sense that it can be recast to an efficient algorithm bringing any
witness to an optimal one \cite{optimization}. As such, optimal EWs
remain a crucial tool in entanglement theory.

The above concepts can be recast in terms of positive maps.
Recall that the Choi-Jamio{\l}kowski isomorphism
\cite{Jisom,Choi} (see also Ref. \cite{Paulsen}) establishes the equivalence
between the set ${\cal L}(M_d, M_{d'})$ of linear maps from $M_d$ to $M_{d'}$
and $M_d \otimes M_{d'}$. Within this pairing $\mathcal{P}$ and
$\mathcal{W}$ correspond to the convex cone of positive maps and its
subset of those maps that are not completely positive, respectively.
Accordingly, a positive map is called optimal if its corresponding EW is
optimal. Then, any element of $\mathcal{D}_{\mathrm{sep}}$ is isomorphic to an
entanglement breaking channel \cite{Holevo} (see also \cite{EB}), that is a
completely positive map $\Lambda: M_d \longrightarrow M_{d'}$ such that (i)
$(I\ot\Lambda)[\rho]\in\mathcal{D}_{\mathrm{sep}}$ for all $\rho\in\mathcal{D}$
and (ii) $\Tr[\Lambda(X)]=\Tr X$ for any $X\in M_d$ (trace-preservicity).

In terms of positive maps the above separability
criterion reads \cite{HorodeckiCriterion}: $\rho \in {\cal
D}_{\mathrm{sep}}$ iff $(I\otimes \Lambda)[\rho] \geq 0$ for any positive map
$\Lambda: M_{d_B}\longrightarrow M_{d_A}$. A positive map gives a more powerful
necessary condition for separability than the corresponding EW: the best known
example is given by the transposition map, which detects all
two-qubit and qubit-qutrit entangled states \cite{HorodeckiCriterion}, whereas
the associated EW only some of them. However, the main drawback of
positive maps in comparison to EWs is that they cannot be realized in physical
experiments, since positive but not CP maps do not represent physical processes.
The \textit{structural physical approximation} (SPA) \cite{PH} is a method that
allows one to overcome this problem by mixing a positive map
$\Lambda$ with the completely depolarizing channel defined as $D(X) =
\mathrm{Tr}(X)\mathbbm{1}_d/d$ for $X\in M_d$. Clearly, for any
positive map $\Lambda$ there exists $p_*\in(0,1)$ such that for any
$p\in[0,p_*]$ the linear map $\Lambda(p)=p\Lambda+(1-p)D$
is completely positive, and as such represents a physical process. For the
largest $p=p_*$ for which this is the case,
we call $\Lambda(p_*)$ the SPA of $\Lambda$. Notice that via the
Choi-Jamio\l{}kowski isomorphism, one can translate the notion of SPA to EWs:
given $W\in\mathcal{W}$, the operator
\begin{equation}
W(p_*)=p_*W+(1-p_*)\frac{\mathbbm{1}_{d_Ad_B}}{d_Ad_B}.
\end{equation}
with $p_*=1/(1+d_Ad_B|\lambda_{\min}|)$, where $\lambda_{\min}<0$ is
the minimal eigenvalue of $W$, is called the SPA of $W$.

\section{A method to test optimality of EWs}
\label{sec:1}

Let us now sketch the method developed in Ref. \cite{optimization}, that we will
later exploit to demonstrate that the EW of Refs. \cite{paper,paper2} is not
optimal. To this end, consider a decomposable $W\in\mathcal{W}$ and let
$\epsilon$ be a positive number and $P$ a positive semi-definite operator acting
on $\mathcal{H}_{AB}$ whose support is orthogonal to $\Pi_W$. We assume that
such a nonzero $P$ exists; otherwise, if there is no such $P$, or, equivalently
$\mathrm{span}\Pi_W=\mathcal{H}_{AB}$, the witness is optimal
(see Sec. \ref{intro}). Clearly, for all normalized product
vectors $\ket{e,f}\in\mathcal{H}_{AB}$ satisfying
\begin{equation}\label{war}
\bra{e,f}W\ket{e,f}> \epsilon
\end{equation}
there always exists $\lambda>0$ (e.g., $\lambda< \epsilon/M$ with
$M=\max_{\ket{e,f}}\bra{e,f}P\ket{e,f}$) such that the inequality
(\ref{inequality}) is fulfilled. It is obvious that the same holds for any
$\epsilon>0$, even arbitrarily small, and corresponding product vectors obeying
(\ref{war}). One then needs to check whether (\ref{inequality}) remains
satisfied by those product vectors which do not obey (\ref{war}), i.e., product
vectors for which $\bra{e,f}W\ket{e,f}\leq \epsilon$, where $\epsilon$ can be
considered a free parameter that can be set arbitrarily small. A product vector
$\ket{e,f}$ obeying the latter inequality must then be ``close''
(in the norm induced by the scalar product in $\cal{H}$, i.e.,
$\|\ket{\psi}\|=\sqrt{\langle\psi|\psi\rangle}$) to one of the elements of
$\Pi_W$. To see it in a more explicit way, consider a sequence of product
vectors $\ket{e_n,f_n}$ such that $\langle e_n,f_n|W|e_n,f_n\rangle\leq
\epsilon_n$ with $\epsilon_n\to 0$ as $n\to\infty$. Assume that elements
$\ket{e_n,f_n}$ in the sequence do not converge, in the norm $\|\cdot\|$ to any
element in $\Pi_W$ as $n\to\infty$. However, since the set of normalized vectors
in a finite-dimensional Hilbert space is compact, the sequence $\ket{e_n,f_n}$
contains a subsequence that converges to some nonzero product vector
$\ket{e',f'}$ satisfying $\langle e',f'|W|e',f'\rangle=0$. By
assumption $\ket{e',f'}\notin\Pi_W$, which contradicts the fact that $\Pi_W$
contains all product vectors satisfying the condition in Eq. (\ref{PW}).

Let us now, for simplicity, restrict to the case
$\mathcal{H}=\mathbbm{C}^{3}\ot\mathbbm{C}^{3}$, but the method that follows can
be straightforwardly generalized to any $d_A$ and $d_B$.
It follows from what we have just said that for any product vector
$\ket{e,f}\in\mathbbm{C}^3\ot\mathbbm{C}^3$ such that $\langle
e,f|W|e,f\rangle\leq \epsilon$ there is a normalized $\ket{e_0,f_0}\in \Pi_W$
such that local components of $\ket{e,f}$
can be written as
\begin{equation}\label{pert1}
\ket{e}=\frac{1}{\sqrt{1+|\delta_1|^2+|\delta_2|^2}}(\ket{e_0}+\delta_1
\ket{e_1}+\delta_2\ket{e_2}),
\end{equation}
and
\begin{equation}\label{pert2}
\ket{f}=\frac{1}{\sqrt{1+|\omega_1|^2+|\omega_2|^2}}(\ket{f_0}
+\omega_1\ket{f_1}+\omega_2\ket{f_2}),
\end{equation}
where $\ket{e_i}$ and $\ket{f_i}$ $(i=1,2)$ are normalized vectors
orthogonal to $\ket{e_0}$ and $\ket{f_0}$, respectively, while
$\delta_i$, and $\omega_i$ $(i=1,2)$ are complex
numbers of vanishing absolute values for $\epsilon\to 0$.

By inserting Eqs. (\ref{pert1}) and (\ref{pert2}) into Eq. (\ref{war}), one
arrives at
\begin{equation}\label{series}
\bra{e,f}W\ket{e,f}=\sum_{i=0}^{4}A_i(W),
\end{equation}
where by $A_i(W)$ we have denoted terms of the $i$th (total) degree in
the variables $\delta_j$, and $\omega_j$ $(j=1,2)$. From
the fact that $\ket{e_0,f_0}\in \Pi_{W}$, and that
the matrices $\bra{e_0}W\ket{e_0}=\Tr_A[(\proj{e_0}\ot\mathbbm{1}_3)W]$ and
$\bra{f_0}W\ket{f_0}=\Tr_B[(\mathbbm{1}_3\ot \proj{f_0})W]$ are
positive semi-definite, it follows that
$A_0(W)=A_1(W)=0$ (analogously, $A_0(P)=A_1(P)=0$ for any $P\geq 0$),
and the first non-vanishing term in Eq.
(\ref{series}) can be $A_2(W)$. On the other hand, $A_3(W)$ and
$A_4(W)$ can be made arbitrarily small by appropriately adjusting
$\epsilon$, which also means that $A_2(W)\geq 0$.
It is then clear that if for any element of $\Pi_W$, $A_2(W)$ is always positive
except for the case $\delta_1=\delta_2=\omega_1=\omega_2=0$,
then for $P\geq 0$ one can adjust $\lambda$
such that the condition (\ref{inequality}) is fulfilled for any $\ket{e,f}$.
In this way we have arrived at the sufficient criterion for an EW $W$ to be
nonoptimal: if $\mathrm{span}\Pi_W\varsubsetneq \mathcal{H}_{AB}$ and
$A_2(W)>0$ for any element of $\Pi_W$ unless
$\delta_1=\delta_2=\omega_1=\omega_2=0$, then $W$ is not optimal.

\section{A proof that the decomposable EW of Ref.
\cite{paper} is not optimal}
\label{sec:2}

We can now pass to the decomposable EW introduced by Ha and Kye
\cite{paper,paper2}. It acts on
$\mathcal{H}_{AB}=\mathbbm{C}^3\ot\mathbbm{C}^3$ and
is given by
\begin{eqnarray}\label{wit}
W_{\theta,b}&=&\proj{w_0}^{T_A}+\frac{1}{b}\sum_{i=1}^{3}\proj{w_i}^{T_A}
\nonumber\\
&=&\left(
\begin{array}{ccccccccc}
1 & 0 & 0 & 0 & \mathrm{e}^{\mathbbm{i}\theta} & 0
& 0 & 0 & \mathrm{e}^{-\mathbbm{i}\theta}\\
0 & b & 0 & 1 & 0 & 0 & 0 & 0 & 0 \\
0 & 0 & \frac{1}{b} & 0 & 0 & 0 & 1 & 0 & 0\\
0 & 1 & 0 & \frac{1}{b} & 0 & 0 & 0 & 0 & 0\\
\mathrm{e}^{-\mathbbm{i}\theta} & 0 & 0 & 0 & 1 & 0 & 0 & 0 &
\mathrm{e}^{\mathbbm{i}\theta}\\
0 & 0 & 0 & 0 & 0 & b & 0 & 1 & 0\\
0 & 0 & 1 & 0 & 0 & 0 & b & 0 & 0 \\
0 & 0 & 0 & 0 & 0 & 1 & 0 & \frac{1}{b} & 0 \\
\mathrm{e}^{\mathbbm{i}\theta} & 0 & 0 & 0 &
\mathrm{e}^{-\mathbbm{i}\theta} & 0 & 0 & 0 & 1
\end{array}
\right),
\end{eqnarray}
with $b>0$, $b\neq 1$, $-\pi/3<\theta<\pi/3$, and $\theta\neq 0$.
The four vectors $\ket{w_i}$ read
\begin{eqnarray}
\ket{w_0}&=&\ket{00}+\ket{11}+\ket{22},\\
\ket{w_i}&=&b\ket{i-1,i}+\mathrm{e}^{\mathbbm{i}\theta}\ket{i,i-1}
\quad (i=1,2,3),
\end{eqnarray}
where the summation is modulo three and $\ket{i}$ $(i=0,1,2)$ form the
standard basis in $\mathbbm{C}^3$. It straightforwardly follows
from (\ref{wit}) that this witness has a block-diagonal form
$W_{\theta,b}=\Lambda_{\theta}+(1/b)(P_{1}+P_{2}+P_{3})$, where
$\Lambda_{\theta}$ is given by
\begin{equation}
\Lambda_{\theta}=\left(
\begin{array}{ccc}
1 & \mathrm{e}^{\mathbbm{i}\theta} &
\mathrm{e}^{-\mathbbm{i}\theta}\\
\mathrm{e}^{-\mathbbm{i}\theta} & 1 &
\mathrm{e}^{\mathbbm{i}\theta}\\
\mathrm{e}^{\mathbbm{i}\theta} & \mathrm{e}^{-\mathbbm{i}\theta} &
1
\end{array}
\right)
\end{equation}
and acts on on the subspace spanned by $\ket{ii}$ $(i=0,1,2)$. The
three remaining blocks are rank-one matrices $P_{i}=\proj{\phi_i}$
with $\ket{\phi_{i}}=b\ket{i-1,i}+\ket{i,i-1}$ $(i=1,2,3)$ being
vectors spanning a subspace orthogonal to $\Pi_{W_{\theta,b}}$, with
the latter, as it is indirectly stated in Ref. \cite{paper}, consisting of the
following six vectors
\begin{eqnarray}\label{PWvecs}
\ket{z_1^{(\pm)}}&=&(\pm1, \omega^*,0)\ot (\omega,\mp1,0),\nonumber\\
\ket{z_2^{(\pm)}}&=&(0,\pm1, \omega^*)\ot (0,\omega,\mp1),\nonumber\\
\ket{z_3^{(\pm)}}&=&(\omega^*,0,\pm1)\ot (\mp1,0,\omega),
\end{eqnarray}
where $\omega=\sqrt{b}\,\mathrm{e}^{\mathbbm{i}\theta/2}$.
Let us finally mention that for any $\theta\in(-\pi/3,\pi/3)$ and
$\theta\neq0$,
the witness has exactly one negative eigenvalue which comes from
$\Lambda_{\theta}$, while for $\theta=0$, $\Lambda_{0}$ is an
unnormalized projector onto $\ket{00}+\ket{11}+\ket{22}$, and
therefore $W_{0,b}$ becomes a positive semi-definite matrix for any positive
$b$.

We are now ready to show that $W_{\theta,b}$ is not optimal. First of all,
one notices that $\Pi_{W_{\theta,b}}$ of $W_{\theta,b}$ consists of only six
vectors (\ref{PWvecs}), and therefore $W_{\theta,b}$ does not have the spanning
property (cf. Sec. \ref{sec:1}). This does not imply, nevertheless, that the
witness is not optimal, because examples of optimal decomposable EWs without the
spanning property are known \cite{RAGSML}.

In what follows, exploiting the method sketched in Sec. \ref{sec:1},
we will show that $W_{\theta,b}$ is indeed not optimal.
To this end, consider a particular pair of vectors
from $\Pi_{W_{\theta,b}}$, say $\ket{z_1^{(\pm)}}$. Up to
normalization, their local components read
$\ket{e_0^{(\pm)}}\propto(\pm1,\omega^*,0)$ and
$\ket{f_0^{(\pm)}}\propto(\omega,\mp1,0)$, while vectors
orthogonal to both of them can be taken as
$\ket{e_1^{(\pm)}}\propto(\mp\omega,1,0)$ and
$\ket{e_2^{(\pm)}}=(0,0,1)$, and
$\ket{f_1^{(\pm)}}\propto(\pm1,\omega^*,0)$
and $\ket{f^{(\pm)}_2}=(0,0,1)$. It then follows
from (\ref{pert1}), (\ref{pert2}) and (\ref{series}) that

\begin{eqnarray}\label{A2}
A^{(\pm)}_2(W_{\theta,b})&=&\frac{1}{N_eN_f}
\left[2(|\delta_1|^2+|\omega_1|^2)+\left(b+\frac{1}{b}
-1\right)(|\delta_2|^2+|\omega_2|^2)\right.\nonumber\\
&&\hspace{1.1cm}\left.\pm\frac{4\sqrt{b}}{1+b}\sin\left(\frac{3\theta}{2}
\right)\mathrm { Im }
(\delta_2\omega_2)\right],
%
\end{eqnarray}
where $N_e=1+|\delta_1|^2+|\delta_2|^2$ and
$N_f=1+|\omega_1|^2+|\omega_2|^2$. It is easily provable
that $A^{(\pm)}_2(W_{\theta,b})$ are strictly positive unless
$\delta_1=\delta_2=\omega_1=\omega_2=0$. With the aid of the facts
that $\mathrm{Im}z\leq |z|$ holds for any $z\in\mathbbm{C}$ and
$|\!\sin(3\theta/2)|\leq 1$, we can lower bound them in the
following way
\begin{eqnarray}
A^{(\pm)}_2(W_{\theta,b})&\geq&\frac{1}{N_eN_f}
\left[2(|\delta_1|^2+|\omega_1|^2)+\left(b+\frac{1}{b}
-1\right)(|\delta_2|^2+|\omega_2|^2)\right.\nonumber\\
&&\hspace{1.1cm}\left.-\frac{4\sqrt{b}}{1+b}|\delta_2\omega_2|\right],
\end{eqnarray}
which can later be rewritten as
\begin{eqnarray}\label{dupa}
A^{(\pm)}_2(W_{\theta,b})&\!\!\!\!\geq\!\!\!\!&\frac{1}{N_eN_f}
\left[2(|\delta_1|^2+|\omega_1|^2)+\left(b+\frac{1}{b}
-1\right)(|\delta_2|-|\omega_2|)^2\right.\nonumber\\
&&\left.\hspace{1.1cm}+2\left(b+\frac{1}{b}-1-\frac{2\sqrt{b}}{1+b}
\right)|\delta_2\omega_2|\right]
,
\end{eqnarray}
It is not difficult to convince oneself that $b+1/b-1>1$, and
\begin{equation}
b+\frac{1}{b}-1-\frac{2\sqrt{b}}{1+b} >0
\end{equation}
for any positive $b\neq 1$; therefore the expression in square
brackets in Eq. (\ref{dupa}) is always positive except for the
case when $|\delta_1|=|\delta_2|=|\omega_1|=|\omega_2|=0$.

For the remaining two pairs of vectors in
$\Pi_W$, one exploits the fact that $\ket{z^{(\pm)}_i}=S^{i-1}\ot
S^{i-1}\ket{z_1^{(\pm)}}$ $(i=2,3)$ with $S$ being a unitary operator
such that $S\ket{i}=\ket{i+1}$ $(i=0,1,2)$, where addition is modulo three.
Thus, the vectors orthogonal to the local components of
$\ket{z^{(\pm)}_i}$ can be taken as $S^{i-1}\ket{e_j^{(\pm)}}$
and $S^{i-1}\ket{f_j^{(\pm)}}$ with $j=1,2$ (recall that $\ket{e_j^{(\pm)}}$ and
$\ket{f_j^{(\pm)}}$ are vectors orthogonal to the local components
of $\ket{z_1^{(\pm)}}$; see above for their explicit forms). One finally checks
that
\begin{equation}
S\ot S\,W_{\theta,b}\,S^{\dagger}\ot S^{\dagger}=W_{\theta,b}.
\end{equation}
All this means that $A_2^{(\pm)}(W_{\theta,b})$
for both pairs $\ket{z_i^{(\pm)}}$ ($i=2,3$) is given by the same formula
(\ref{A2}). As a consequence, $W_{\theta,b}$ is not optimal for any
positive $b\neq 1$ and $| \theta| \in (0, \pi/3)$, and hence cannot
serve as a counterexample to the conjecture.

\begin{figure}[]
\includegraphics[width=0.49\columnwidth]{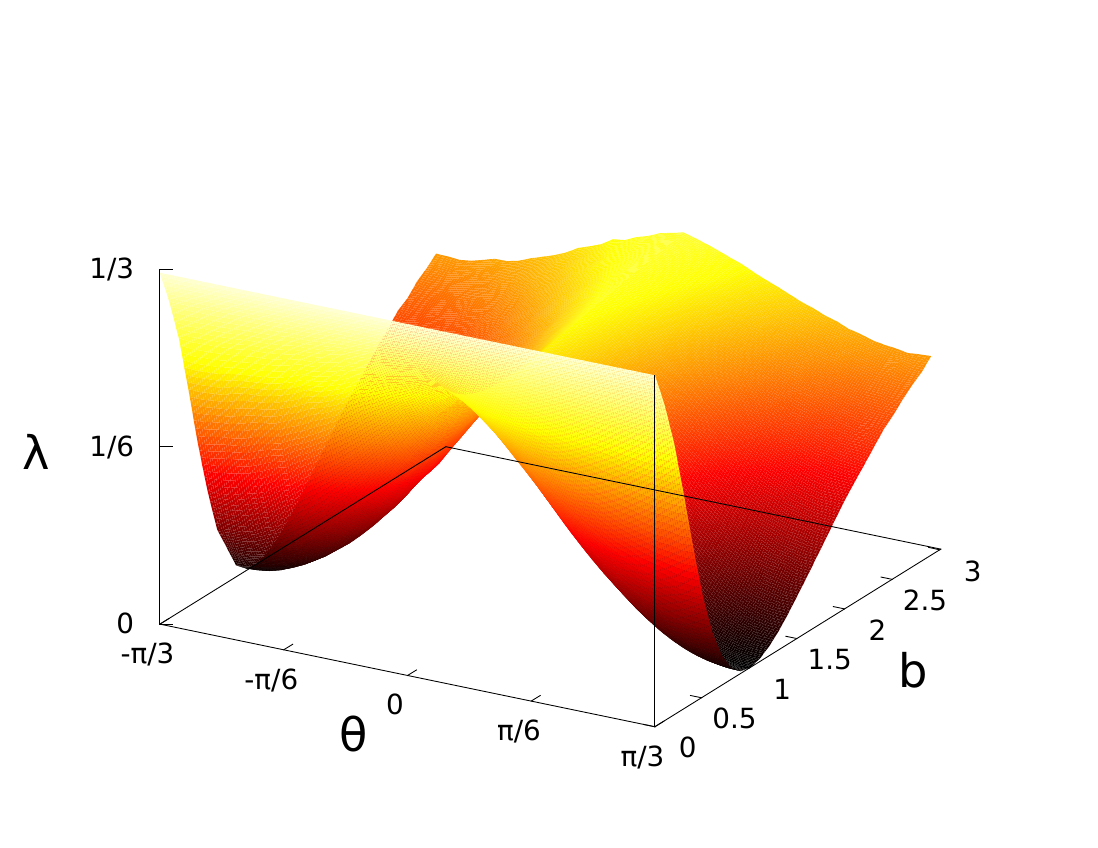}
\includegraphics[width=0.49\columnwidth]{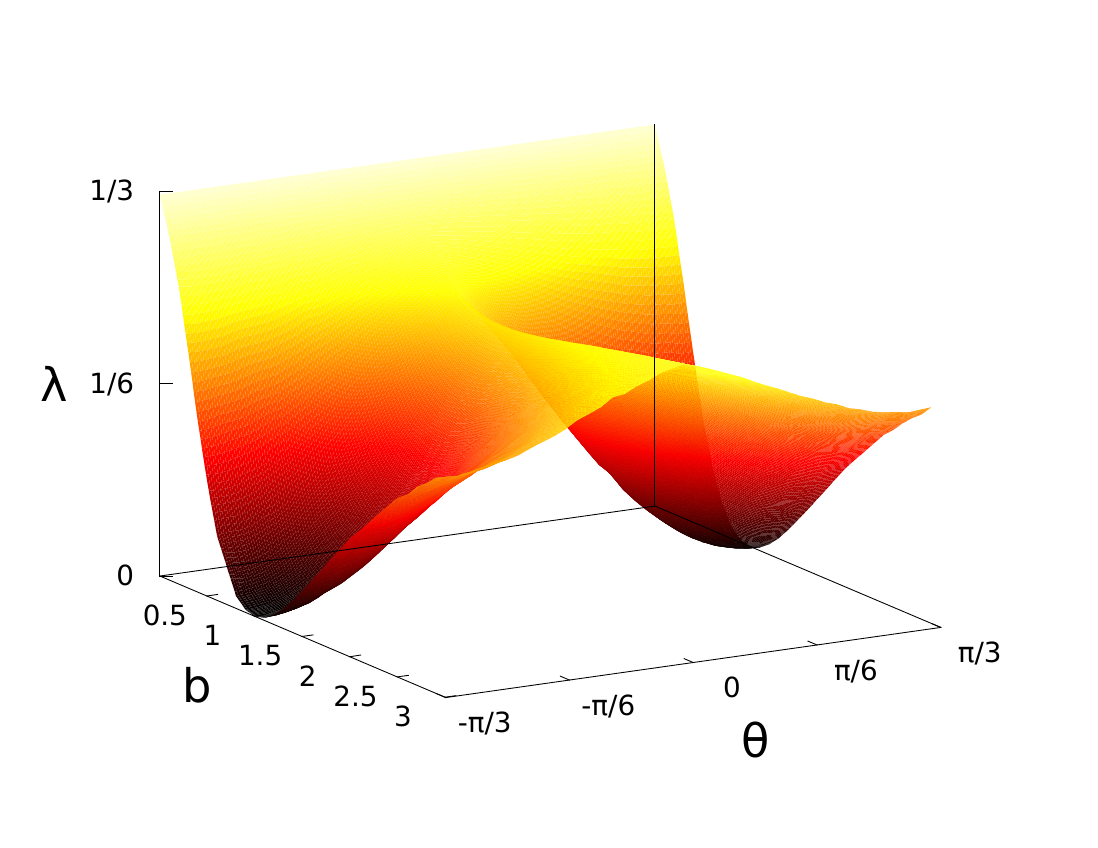}
\caption{As a function of $\theta\in[-\pi/3,\pi/3]$ and
$b\in(0,3)$, it is presented the numerically determined maximal
value of $\lambda$ such that $(\widetilde{W}_{\theta,b}-\lambda
\widetilde{P}_1)/(1-\lambda)$ is an entanglement witness. Here
$\widetilde{W}_{\theta,b}$ and $\widetilde{P}_1$ denote the
normalized witness $W_{\theta,b}$ and the rank-one operator $P_1$.
The visible maximum for $\theta=0$ is a consequence of the fact
that $W_{0,b}$ is a positive matrix for any allowed $b$, and hence
one can subtract the whole block $(1/b)P_1$ from it.
}\label{fig:1}
\end{figure}

With the aid of the procedure described in Ref. \cite{Leinaas} we
have also numerically studied optimality of the entanglement witness
(\ref{wit}). Since the optimization over product vectors may lead to different
local minima, several initial conditions, uniformly and randomly chosen
have been considered, and this procedure was repeated at least
$10^3$ times for each choice of $\theta$ and $b$ (the increment step was taken
as $0.05$ and $0.1$, respectively). It clearly follows from Fig. \ref{fig:1}
that the obtained results fully support the above proof.
As a positive operator to be subtracted from the witness we took
$\widetilde{P}_1=P_1/(1+b^2)$ (see above for the definition of
$P_1$). Then, as a function of $\theta\in[-\pi/3,\pi/3]$ and
$b\in(0,3)$ we determined the maximal value of $\lambda$ such that
$\widetilde{W}(\lambda)=(\widetilde{W}_{\theta,b}-\lambda
\widetilde{P}_1)/(1-\lambda)$ is an entanglement witness (see Fig.
\ref{fig:1}), where $\widetilde{W}_{\theta,b}$ is a normalized
version of $W_{\theta,b}$. It should be noticed that $W_{0,b}$ is a positive
matrix and therefore one can remove the whole block $(1/b)P_1$ from it. This
corresponds to the fact that for any fixed $b$, the function on Fig. \ref{fig:1}
achieves its maximum for $\theta=0$.

\section{Concluding remarks}

We have shown in this note, exploiting the method formulated in Ref.
\cite{optimization}, that the decomposable entanglement witness introduced in
Refs. \cite{paper,paper2} is not optimal. This in particular means that, 
contrary to the claim of \cite{paper}, it does not disprove the SPA conjecture
in the decomposable case. Nevertheless, a proper counterexample has recently
been found by other researchers \cite{Gniewko3}. This together with an analogous
counterexample in the indecomposable case provided by Ha and Kye \cite{HaKye1},
fully disproves the SPA conjecture. 

Still the fact that SPAs of some optimal positive maps are entanglement
breaking is interesting because one of its main advantages is that
the physical implementation of the resulting completely positive maps can be
simplified to a measure \& prepare scheme (see \cite{Korbicz}). Therefore, it
is reasonable to ask as to whether the notion of SPA can be modified so that the
resulting completely map ``best approximating'' a given positive map is
entanglement breaking \cite{we,Myrheim}. And, actually, it was already shown in
Ref. \cite{we} that for any optimal decomposable map there exists an
entanglement breaking channel (not necessarily the fully depolarizing one) such
that their mixture with the smallest possible mixing parameter necessary to
obtain a completely positive map, is another entanglement breaking channel.
In the language of entanglement witnesses this means that for any optimal
decomposable witness there is a separable state such that their combination
with the smallest ``weight'' necessary to turn the obtained
operator into a positive one, is a separable state. And, the obtained
approximation is the best in the sense that the resulting operator
lies on the boundary of the convex set of positive operators.
In the indecomposable case, however, the question is left open.

\ack{Discussions with J. Stasi\'nska are acknowledged.
This work is supported by the ERC Grant QUAGATUA, EU projects AQUTE and SIQS,
and Spanish TOQATA. R. A. acknowledges support from the Spanish MINECO through
the Juan de la Cierva program. J. B. acknowledges National Research Foundation
and Ministry of Education of Singapore.}

\end{document}